\documentclass[conference]{IEEEtran}
\ifCLASSINFOpdf
\else
\fi

\usepackage{url}
\usepackage{graphicx}
\usepackage{subfigure}
\usepackage{mdwlist}

\begin{document}
%
\title{Object-Oriented Sokoban Solver: A Serious Game Project for OOAD and AI Education}

\author{\IEEEauthorblockN{Zheng Li}
\IEEEauthorblockA{School of Computer Science\\
ANU and NICTA\\
Canberra, Australia\\
zheng.li@nicta.com.au}
\and
\IEEEauthorblockN{Liam O'Brien}
\IEEEauthorblockA{ICT Innovation and Services\\
Geoscience Australia\\
Canberra, Australia\\
liamob99@hotmail.com}
\and
\IEEEauthorblockN{Shayne Flint}
\IEEEauthorblockA{School of Computer Science\\
Australian National University\\
Canberra, Australia\\
shayne.flint@anu.edu.au}
\and
\IEEEauthorblockN{Ramesh~Sankaranarayana}
\IEEEauthorblockA{School of Computer Science\\
Australian National University\\
Canberra, Australia\\
ramesh@cs.anu.edu.au}}

%


\maketitle

\begin{abstract}
Serious games are beneficial for education in various computer science areas. Numerous works have reported the experiences of using games (not only playing but also development) in teaching and learning. Considering it could be difficult for teachers/students to prepare/develop a game from scratch during one semester, assistant educational materials would be crucial in the corresponding courses. Unfortunately, the literature shows that not many materials from educational game projects are shared. To help different educators identify suitable courseware and help students implement game development, it is worth further investigating and accumulating the educational resources from individual game projects. Following such an idea, this paper proposes a game development project of an object-oriented Sokoban solver, and exposes relevant educational materials. The documented system design can be viewed as a ready-to-use resource for education in object-oriented analysis and design (OOAD), while the Sokoban solver itself may be used as an assignment platform for teaching artificial intelligence (AI). Further documentation, platform, and APIs will be realized and shared in the future to facilitate others' educational activities. Overall, this work is supposed to inspire and encourage other researchers and educators to post available materials of more game projects for the purpose of sharing and reuse.
\end{abstract}


%
\IEEEpeerreviewmaketitle

\section{Introduction}
%
%

%
%
%
%
Serious games usually refer to computer games designed for non-entertainment purposes, which have been identified as being useful for education in almost every area of computer science \cite{Jones_2000,Wong_Zink_2010}. A wide range of computer course topics can be represented by game-related knowledge, such as programming skills, algorithm and data structure, object-oriented analysis and design (OOAD), artificial intelligence (AI), etc. Moreover, employing educational game projects in courses may offer various additional benefits ranging from increasing students' learning interests to creating interactive learning environments \cite{Jones_2000,Longstreet_Cooper_2011,Ludi_2011,Wong_Zink_2010}. As such, universities have started investigating how to use games to teach computer science and software engineering \cite{Longstreet_Cooper_2011}.

Unlike most studies that emphasize game playing, in our school, the processes of developing serious games seem to be more valuable and suitable for some relevant courses. When preparing the course contents, unfortunately, the lack of resources becomes one of the concerns to realize the full potential of educational game development. Although many studies have shared the experiences of using games in education \cite{Ludi_2011,Wong_Zink_2010}, few works reported detailed materials of the corresponding game projects. In fact, the assistant resources would be crucial and beneficial for both teachers and students in educational game development. For example, teachers need documents and technical supports to familiarize themselves with the games, and then identify the relevance of a game to the curricula \cite{Barbosa_Silva_2011}; while students may need program modules, libraries and even semi-finished products to focus on the most relevant learning activities, rather than developing a game from scratch. Thus, the lack of open resources could also be one of the reasons for such a phenomenon: although applying games to education has gained acceptance widely, the real employment of serious games in schools remains limited \cite{Barbosa_Silva_2011,Susi_Johannesson_2007}.

\begin{figure}[!t]
\centering
\includegraphics[width=8.5cm]{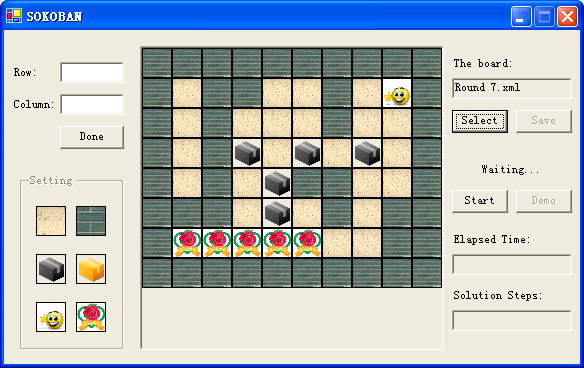}
\caption{\label{fig>PicSokoban}Interface of the proposed object-oriented Sokoban solver with a sample game round.}
\end{figure}

Therefore, to facilitate applying serious game development to education, a promising strategy would be accumulating and opening available materials to enrich the game-based educational resources. 
As an initial step to sharing educational materials, this paper introduces a potential game project - namely object-oriented Sokoban solver - for both OOAD education and AI education. On the one hand, this project provides an excellent example of object-oriented environment, which confirms the generic advantages of courses on computer games \cite{Jones_2000}. Based on the object-oriented analysis, the Sokoban solver can be designed to demonstrate most of the important object-oriented concepts and typical relationships between classes/objects. On the other hand, this project offers an ideal platform for teaching \textit{search} in the AI domain. Students may be asked to implement different path-finding algorithms to move single or multiple boxes on a grid with or without obstacles (cf.~Fig.~\ref{fig>PicSokoban}). Different search processes and the generated paths can be played as animations for intuitive comprehension, comparison, and even competition. Furthermore, since Sokoban solving still remains an interesting research topic \cite{Demaret_Lishout_2008}, this project could encourage students to continue and promote higher levels of learning.

The remainder of this paper is organized as follows. 
Section \ref{II} specifically maps the design of this Sokoban solver with important object-oriented concepts and relationships, which is the main contribution of this paper. Since different AI strategies of Sokoban solving have been well-documented in the related literature, this work only briefly summarizes several typical techniques and illustrate a simple case study in Section \ref{III}. Conclusions and future work are discussed in Section \ref{IV}.

\section{Sokoban Solver as a Serious Game Project for OOAD Education}
\label{II}
\begin{figure*}[!t]
\centering
\includegraphics[width=14cm]{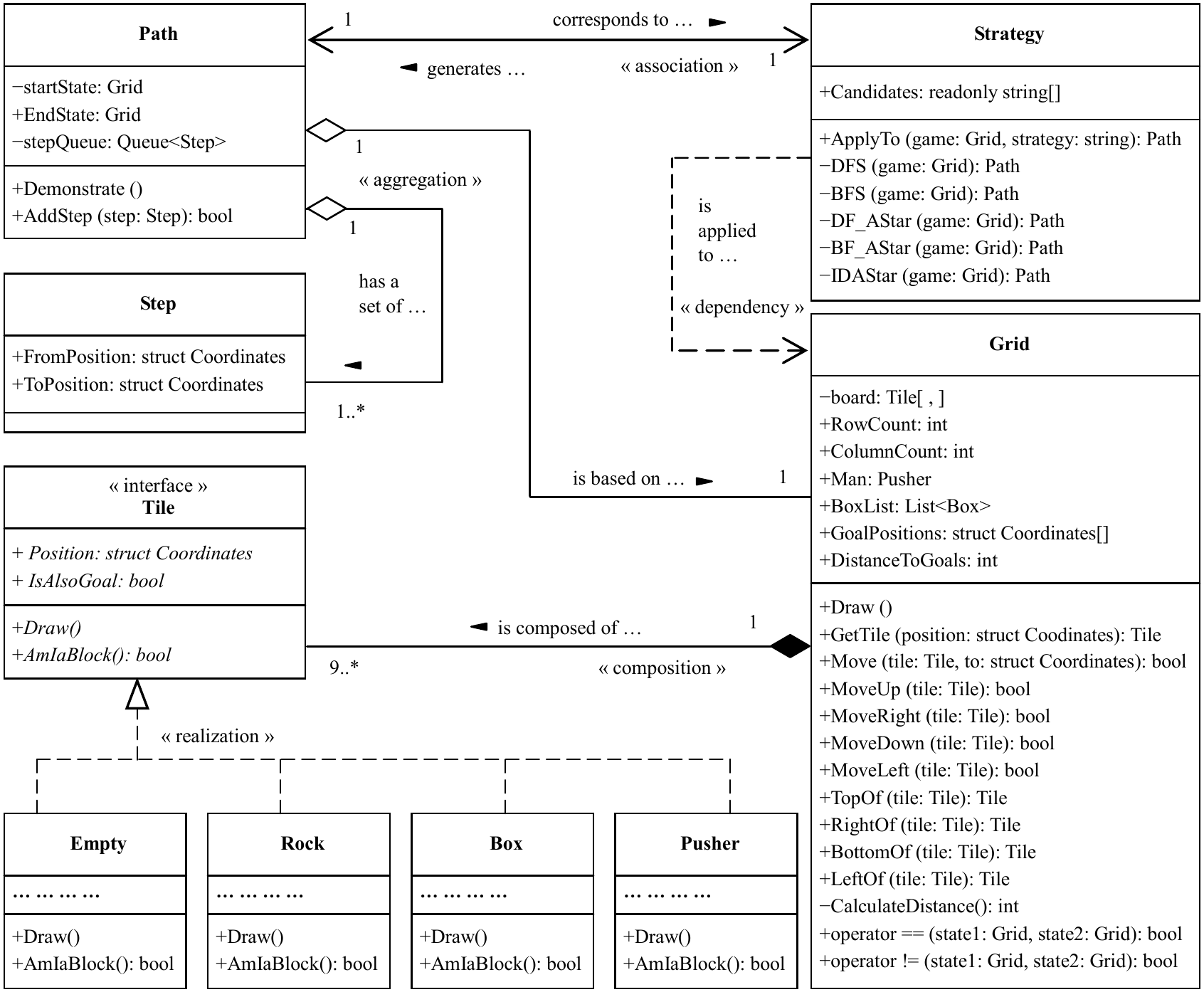}
\caption{\label{fig>PicClass}Class diagram for object-oriented Sokoban solver.}
\end{figure*}

Object orientation has become one of the primary paradiagms in undergraduate education \cite{Chen_Cheng_2007}. Unfortunately, the deep object orientation concept often get lost by beginning learners \cite{Berndtsson_2005}. A survey shows that around $30\%$ of computer science students may not understand the basics of OOAD \cite{Wei_Moritz_2005}. Since computer games are natural simulations that involve a large number of interacting objects \cite{Chen_Cheng_2007}, Sokoban was selected as an attracting sample to vividly demonstrate important object orientation concepts and class/object relationships ranging from \textit{abstraction} to \textit{polymorphism}. When applying Unified Modeling Language (UML) to interpret these concepts and relationships, to save space, this paper only illustrates the class diagram, as shown in Fig.~\ref{fig>PicClass}.

\subsection{Abstraction}
The central issue in OOAD for a given problem is to identify the right abstractions to discover objects. In the context of OOAD, ``an abstraction denotes the essential characteristics of an object that distinguish it from all other kinds of object and thus provide crisply defined conceptual boundaries, relative to the perspective of the viewer" \cite{Booch_Maksimchuk_2007}. In this case, the Sokoban solver acts as an automatic game player moving boxes to the goal squares on a grid. Correspondingly, there are two types of abstractions in this domain, namely Entity abstraction and Action abstraction \cite{Booch_Maksimchuk_2007}. The Entity abstraction refers to the Sokoban environment: a particular grid composed of a set of tiles; the Action abstraction indicates the game playing process: an AI strategy finally discovers a solution that is composed of a sequence of steps. Consequently, five classes of objects can be determined: Tile, Grid, Step, Path, and Strategy. In particular, the Tile object can be further distinguished between Empty, Rock, Box, Goal, and Pusher. Considering the objects of Box/Pusher and Goal can overlap each other in the game, the Goal is finally removed while merged with other Tile objects through a boolean attribute \textit{IsAlsoGoal}.  

\subsection{Encapsulation}
As one of the significant characteristics of OOAD, encapsulation complements the aforementioned concept abstraction. Abstraction reflects the observation of objects from outside, while encapsulation implies the implementation of objects from inside. Moreover, encapsulation requires hiding information of objects as much as possible to maximize the objects' independence, so as to make the generated software more robust and easier to maintain \cite{Wampler_2002}. As such, by using the programming mechanism ``classes" to realize encapsulation, Grid encapsules its size and monitors where the movable tiles are; static Tiles keep their positions and types, while movable Tiles would have behaviors like moving up/right/down/left or even moving to a different position; Step encapsules its ID and records a move of a particular box; Path comprises a set of useful steps and tells how to shift from one step to another; Strategy hides the implementation of an AI algorithm.

\subsection{Class Hierarchies}
When it comes to creating object-oriented programs, one of the most important aspect is to arrange classes into hierarchies to reflect the relationships between different objects \cite{Wampler_2002}. For this object-oriented Sokoban solver, five types of relationships can be identified: association, composition, dependency, inheritance, and aggregation, which are specified respectively in the following subsections.

\subsubsection{Association}
Association denotes the most general relationships among classes. Although this class relationship is the most semantically weak, association identification is particularly helpful for analyzing and early designing a software system \cite{Booch_Maksimchuk_2007}. As the design and implementation work continues, the weak associations can be gradually refined into other concrete relationships. For example, according to the above-mentioned abstraction, Grid and Strategy has an association with many-to-many multiplicities: one Sokoban grid can be solved by using different strategies, while one strategy can be applied to different grids. However, since various search algorithms can be pre-implemented in the Strategy class, this association may be further turned into a dependency relationship, as described below.

\subsubsection{Dependency}
One class depends on another if it uses concrete objects of the other class \cite{Horstmann_2006}. When implementing the proposed Sokoban solver, a dependency relationship can be found between Strategy and Grid. In reality, the AI strategy cannot be applied without a particular instance of Grid. Note that the Strategy class does not need to maintain a Grid attribute, but uses Grid as a parameter variable for its \textit{ApplyTo()} method. In fact, this is also the characteristic of dependency in OOAD.

\subsubsection{Aggregation}
Aggregation can be viewed as a special case of dependency \cite{Horstmann_2006}. One class aggregates other classes if it treats the others as its parts, which implies that the former must be aware of the existence of the other classes. In general, it is possible to use ``has-a" or ``part-of" to informally describe the aggregation relationship \cite{Wampler_2002}. For example, based on the encapsulation analysis, one Solution should have a Step queue and a related Grid. Thus, it is natural to identify the aggregation between Solution and Step/Grid. Different from the implementation of dependency, aggregation here is represented by employing Step and Grid as attributes of the Solution class. 

\subsubsection{Composition}
Composition is a stronger form of aggregation \cite{Horstmann_2006}. The parts cannot exist independently of their host in a composition relationship. While constructing the interface of Sokoban, composition takes place between Grid and Tiles. Student may be suggested to semantically distinguish between composition and aggregation. In this case, it does not make sense to keep particular Tile instances if their constructed Grid instance is destroyed. On the contrary, even without Solution in the previous case, Grid is still useful for the game displaying; the Step queue may still be valid for other instances of Grid, although it could be not able to drive boxes to the targets.

\subsubsection{Inheritance}
Inheritance is one of the fundamental features of OOAD \cite{Wampler_2002}, which expresses the generalization/specialization relationships between classes \cite{Booch_Maksimchuk_2007}. A class inherits from another if all objects of the former class are special cases of objects of the latter \cite{Horstmann_2006}. In the abstraction of Sokoban solver, five different concrete Tile objects on a Grid have been identified, namely Empty, Rock, Box, Goal, and Pusher. Then, it is natural to design five specialized classes inherited from a general Tile class. As such, the common behaviors of the general Tile class will be reused, while additional responsibilities can still be supplemented to the specialized classes. In particular, for the educational purpose, the general Tile classes can be further designed as an \textit{interface}. An \textit{interface} provides the outside view of the related class. It describes the class's behaviors while does not contain implementations. For example, the \textit{interface} Tile may declare an \textit{AmIaBlock()} method to force every specialized class to supply help for recognizing obstacles. 

\subsection{Polymorphism}
Polymorphism is closely related to the aforementioned inheritance relationship between classes. When a set of subclasses inherit from a superclass, each of the subclasses can extend properties and override behaviors of the superclass. Through dynamic binding at runtime \cite{Wampler_2002}, polymorphism supplies the ability to select appropriate behaviors for different subclass objects according to their actual types \cite{Horstmann_2006}. As for the previous case of different classes inheriting from Tile, Rock and Box are implemented to return \textit{TRUE} for their \textit{AmIaBlock()} methods, while the others return \textit{FALSE}. Then, the interface Tile's \textit{AmIaBlock()} method may be directly employed during programming, while the correct methods will be called for path finding and deadlock identification \cite{Cazenave_Jouandeau_2010} when solving a Sokoban game. As such, polymorphism essentially promotes loose coupling \cite{Booch_Maksimchuk_2007} and facilitates programming.


\section{Sokoban Solver as a Serious Game Project for AI Education}
\label{III}
The use of games in AI education has also been suggested to attract more students into computing \cite{Wong_Zink_2010}. In this case, the outcome of the previous work would offer a Sokoban game courseware, so that the students can direclty focus on the AI strategies without having to start from scratch. The existing AI strategies of Sokoban solving can be distinguished between single-agent search and multi-agent search \cite{Demaret_Lishout_2008}. The single-agent search further covers a bunch of techniques, such as Depth-First Search (DFS), Breadth-First Search (BFS), A* based either on DFS or on BFS, Iterative-Deepening A* (IDA*), Genetic Algorithm, etc.~\cite{Dorst_Gerontini_2011} 

\begin{figure}[!b]
\centering
\includegraphics[width=7.45cm]{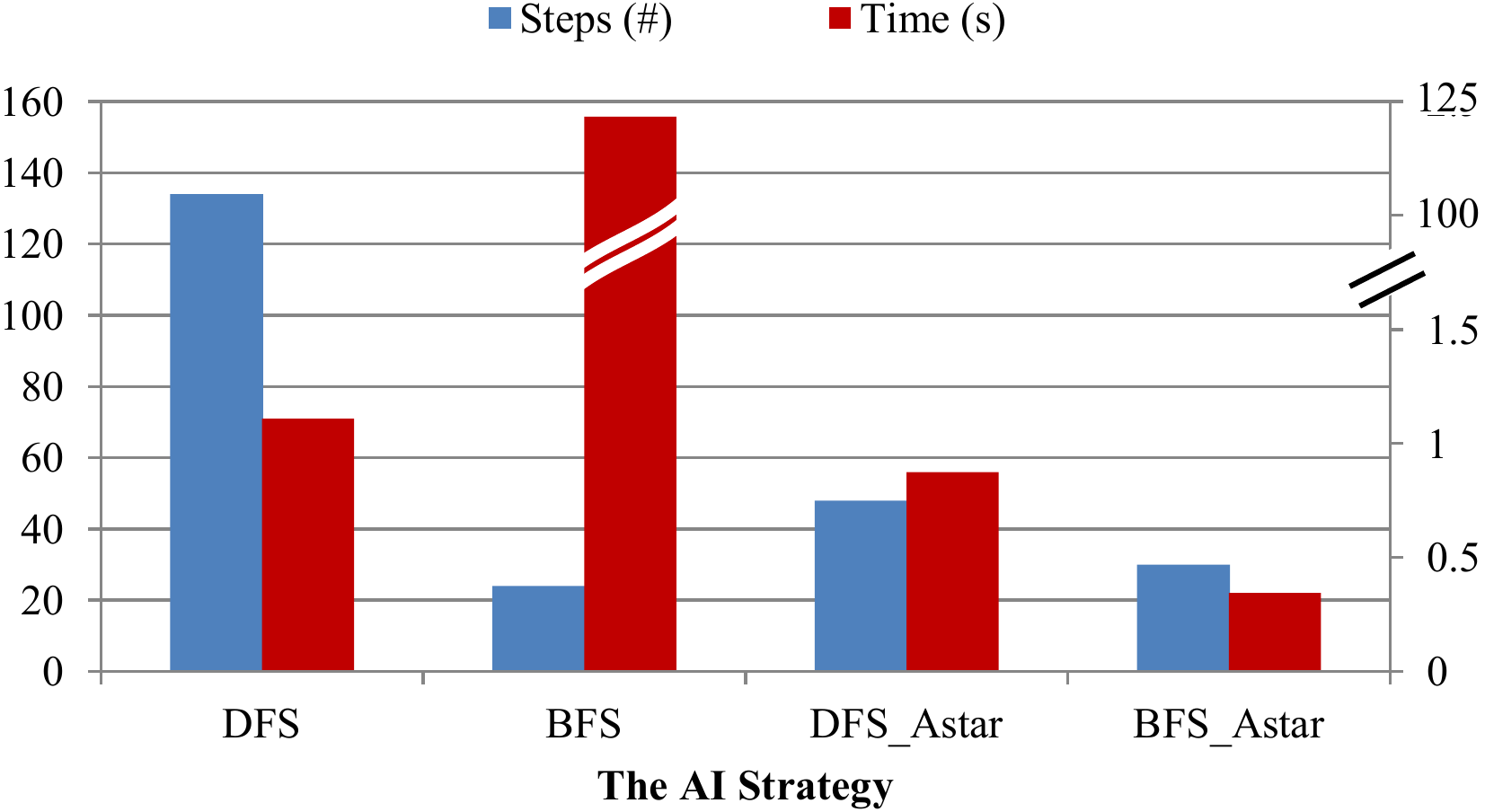}
\caption{\label{fig>PicResult}Comparison between different AI strategies for solving the previous Sokoban puzzle (cf.~Fig.~\ref{fig>PicSokoban}).}
\end{figure}

However, Sokoban solving is an NP-hard problem \cite{Dor_Zwick_1999}, and no solver is able to solve all the benchmark puzzles at present \cite{Demaret_Lishout_2008}. Even for the simple cases, there is no one-size-fits-all strategy to deal with the enormous variety of the game \cite{Dorst_Gerontini_2011}. For example, when running this object-oriented Sokoban solver with four techniques respectively, solving the previous sample puzzle requires different solution steps and elapsed time, as listed in Table \ref{tbl>GameResult}. The result comparison (cf.~Fig.~\ref{fig>PicResult}) shows that: although BFS found the shortest-path solution, it was significantly time-consuming; DFS identified the longest-path solution; while BFS-based A* performed generally better. Nevertheless, the result varied much when testing other simple puzzles, and DFS-based A* gave better overall performance in most of the tested cases.
Note that, considering the calculation of real heuristic functions could be as difficult as $O(n^3)$ for a problem of $n$ boxes \cite{Demaret_Lishout_2008}, a pre-paired sequence of boxes and goals is used to simulate their Manhattan distance.

\begin{table}[!t]\footnotesize
\renewcommand{\arraystretch}{1.3}
\centering
\caption{\label{tbl>GameResult}Result of Solving the Previous Sokoban Puzzle (cf.~Fig.~\ref{fig>PicSokoban})}
\begin{tabular}{|l|c|c|}
\hline
AI Strategy & Solution Steps (\#) & Elapsed Time (s)\\
\hline
DFS & 134& 1.11 \\
\hline
BFS & 24& 121.69\\
\hline
DF\_AStar & 48 & 0.87 \\
\hline
BF\_AStar & 30 & 0.34\\

\hline
\end{tabular}
\end{table}

When it comes to the AI education, it is clear that Sokoban solving can vividly illustrate the effects and characteristics of different strategies, for instance, the usage of heuristics or not. Therefore, this proposed Sokoban solver may either be adopted as a courseware for lecturers to interpret different path-finding techniques, or as an assignment project for students to get familiar with particular AI algorithms.



\section{Conclusions and Future Work}
\label{IV}
Employing serious games in computer science education has been propagated by many researchers through their experience reports. The processes of developing serious games seem to be particularly helpful and valuable for the educational activities in our school. However, a controversial issue is that the practices of educational games are not widespread enough \cite{Annetta_2008}, and we also experienced difficulty when designing and preparing the relevant course contents. It has been suggested that the lack of reusable resources could be one of the obstacles of game-based education \cite{Barbosa_Silva_2011}. Therefore, it would be useful and crucial to clearly document and then share the existing game projects to facilitate future educational activities. This paper proposes a development project of an object-oriented Sokoban solver for both OOAD education and AI education. Such an effort can be viewed as an initial approach to accumulating ready-to-use resources of serious game projects. The future work is then able to be unfolded along two directions: first, we will gradually enrich the documentation of this project, and build a semi-finished Sokoban solver as scaffold to facilitate students' development activities; second, we will employ this educational project in our classes to validate this proposal and share our experiences.

\end{document}